\documentclass[preprint,12pt]{elsarticle}
\usepackage[all]{xy}
\usepackage{amssymb, amsmath}
\usepackage{accents}

\def\T{\mathcal T}

\def\D{\mathcal D}
\def\OO{\mathcal O}

\def\J{\mathcal J}
\def\P{\mathcal P}

\def\G{\overline{\mathcal{G}}}

\usepackage{graphicx}
\usepackage[caption=false]{subfig}

\journal{arXiv}

\begin{document}

\begin{frontmatter}

\title{Solution of the classical Yang--Baxter equation with an exotic symmetry, and integrability
of a multi-species boson tunneling model}
\author{Jon Links\\ }
\address{School of Mathematics and Physics, \\The University of Queensland, Brisbane, QLD 4072, 
Australia \\
email: jrl@maths.uq.edu.au}
\begin{abstract}
Solutions of the classical Yang-Baxter equation provide a systematic method to construct integrable quantum systems in an algebraic manner. A Lie algebra can be associated with any solution of the classical Yang--Baxter equation, from which commuting transfer matrices may be constructed. This procedure is reviewed, specifically for solutions without skew-symmetry. A particular solution with an exotic symmetry is identified, which is not obtained as a limiting expansion  of the usual Yang--Baxter equation. This solution facilitates the construction of commuting transfer matrices which will be used to establish the integrability of a multi-species boson tunneling model. The model generalises the well-known two-site Bose-Hubbard model, to which it reduces in the one-species limit. Due to the lack of an apparent reference state, application of the algebraic Bethe Ansatz to solve the model is prohibitive. Instead, the Bethe Ansatz solution is obtained by the use of operator identities and tensor product decompositions.
\end{abstract}



\end{frontmatter}

\newpage 
\section{Introduction}

\noindent 
The Yang--Baxter equation is a cornerstone in the theory of integrable quantum systems. It facilitates the construction of commuting transfer matrices which generate sets of conserved operators. A limiting procedure may be performed to obtain the {\it classical} Yang--Baxter equation \cite{bd82}. Despite the name, this equation also has a long history in the study of integrable {\it quantum} systems \cite{eks93,j89,j89a}. In some cases these models have been formulated through the notion of a {\it Gaudin algebra} \cite{bt07,g76,osdr05}, which turns out to be an essentially equivalent approach up to generalisation of the classical Yang-Baxter equation \cite{bdp05}. Below, the use of a non-standard form of the classical Yang--Baxter equation will be employed, which affords solutions that do not have the property of being {\it skew-symmetric}. A particular case will be identified within a class of solutions of this equation, and analysed in detail.       

The motivation for this study arises from a recent construction of an integrable generalisation, and Bethe Ansatz solution, of the $p+ip$-pairing Hamiltonian \cite{lil16}. 
With the benefit of some hindsight, it will be shown below that this result can be re-derived in a more direct manner. Not only does this provide a clearer understanding of the integrability of the model derived in \cite{lil16}, it also opens up new avenues of applications. As an example, an integrable, multi-species, boson tunneling model will be derived. 
Following the experimental realisation of a two-species Bose-Einstein condensate in 2008 \cite{tbdcmi08}, there has been considerable effort to study such systems. Tunneling models for two-species systems were quickly formulated \cite{cno11,qtf10,sbnhec09,sp09}, and the general area of research continues to attract significant activity     
\cite{cnpv15, cvct15, lcd14,wpc16,ws15}. Having an integrable model for multi-species tunneling offers an opportunity for a different perspective on tunneling phenomena.


The integrable model which will be derived below is described in terms of $2L$ mutually commuting sets of canonical boson operators, representing $2L$ degrees of freedom for the system.  These operators are labeled by two indices, $j=1,...,L$ denoting the boson species, and $X=A,B$ denoting the two potential wells between which tunneling occurs. The operators satisfy the relations 
\begin{align*}
[b^{\phantom{\dagger}}_{j,X}, \,b^{\phantom{\dagger}}_{k,Y}]&=0, \\
[b^{{\dagger}}_{j,X}, \,b^\dagger_{k,Y}]&=0, \\
[b^{\phantom{\dagger}}_{j,X}, \,b^\dagger_{k,Y}]&=\delta_{jk}\delta_{XY}I.
\end{align*}  
Number operators for each species in each well are given by 
\begin{align*}
N_{j,X}=b^\dagger_{j,X}b^{\phantom{\dagger}}_{j,X}.
\end{align*}
Set
\begin{align*}
N_{X}&=\sum_{j=1}^L N_{j,X}, \\
N_j&=N_{j,A}+N_{j,B}.
\end{align*}
The Hamiltonian is given by
\begin{align}
H=U(N_A-N_B)^2+\mu(N_A-N_B)+\sum_{j=1}^L {\mathcal E}_j \left(b^\dagger_{j,A}b^{\phantom{\dagger}}_{j,B}
+b^\dagger_{j,B}b^{\phantom{\dagger}}_{j,A}\right).
\label{ham}
\end{align}
The coupling parameter $U$ characterising the inter-boson interactions is positive for repulsive interactions and negative for attractive interactions. The well bias, such as that due to the presence of an applied external field, is parametrised by $\mu$. The parameters ${\mathcal E}_j$ are the amplitudes associated with each species for tunneling between the wells. The signs of ${\mathcal E}_j$ are not important, as they can be changed by a unitary transformation $b_{j,X} \mapsto -b_{j,X}$.    
There are no constraints imposed on any of the coupling parameters, they are all free, real variables.

When $L=1$ (\ref{ham}) reduces to the two-site Bose--Hubbard model, which endures as the subject of widespread investigation, e.g. \cite{b16,cdp16,gks14,safr15,css14,wzhql16}. Thus (\ref{ham}) represents an $L$-species generalisation. This Hamiltonian commutes with each of the operators $N_j$, $j=1,..,L$, indicating that the number of particles for each species is conserved. The main objective below is to construct an additional $L$ conserved operators, establishing that the model is integrable.

In Sect. 2 a general Lie-algebraic approach will be taken to construct a family of commuting transfer matrices. It will be seen that the classical Yang-Baxter equation emerges in a natural way, in a form that does not assume skew-symmetry. This form has been promoted by Skrypnyk in a series of works \cite{s06,s07,s07a,s09,s10,s14,s15}. Relationship to the usual form of the Yang--Baxter equation will be discussed. A study of some provisional elements of the associated representation theory will lead to the identification of a distinguished solution of the classical Yang--Baxter equation with an exotic symmetry, in a sense which will be defined. 
In Sect 3. attention turns to the construction of the model, and its solution. Due to the lack of an apparent reference state, application of the algebraic Bethe Ansatz to solve the model in general is prohibitive. Instead the Bethe Ansatz solution will first be derived in the limiting, but non-trivial, case when $N_j=1$ for all $j=1,...,L$.  Based on this result the general solution will be obtained through a tensor product construction. 
Concluding remarks are offered in Sect. 4.

\section{Commuting transfer matrices from the classical Yang-Baxter equation} \label{}
\noindent 
Let $\{e^j_k:j,k=1,...,n\}$ denote the standard basis elements for ${\rm End}({\mathbb C}^n)$. Introduce an abstract, infinite-dimensional complex Lie algebra ${\mathcal G}$ with generating functions 
$$\{T^j_k(u)=\sum_{l=-\infty}^{\infty}t^j_k[l]u^l:u\in{\mathbb C}, j,k=1,...,n\}$$ and set 
\begin{align*}
\T(u)&=\sum_{j,k=1}^n T^j_k(u)e_j^k.
\end{align*}
For $u,\,v\in{\mathbb C}$, let $r(u,v)\in\,{\rm End}({\mathbb C}^n\otimes {\mathbb C}^n)$,  which is expressible as  
\begin{align*}
r(u,v)&=\sum_{j,k,l,p=1}^n r^{jl}_{kp}(u,v)e^k_j\otimes e^p_l.
\end{align*}
The symmetry properties
\begin{align}
[r_{12}(u,v),\,r_{21}(v,u)]&=0, \label{comm1}\\
[r_{12}(u,v)^{{\rm t}_2},\,r_{21}(v,u)^{{\rm t}_2}]&=0,  \label{comm2}
\end{align}
where ${\rm t}_2$ denotes the partial transpose taken in the second component of the tensor product, are assumed. 
These assumptions provide for a different approach than those followed in \cite{s06,s07,s07a,s09,s10,s14,s15}. 
The commutation relations imposed on ${\mathcal G}$ are compactly given by
\begin{align}
\left[ \T_1(u),\,\T_2(v)\right]=\left[ \T_2(v),\,r_{12}(u,v)   \right]-\left[\T_1(u),\,r_{21}(v,u)  \right]
\label{comm}
\end{align}
where the subscripts denote the actions on tensor components of ${\mathbb C}^n \otimes {\mathbb C}^n$. 
Assuming that the Jacobi identity holds, such that an associative enveloping algebra can be constructed in the usual way,  
it follows from the commutation relations (\ref{comm}) that the transfer matrix
\begin{align}
t(u)={\rm tr}\left( (\T(u))^2 \right)=\sum_{j,k=1}^nT^j_k(u)T^k_j(u)
\label{tm}
\end{align}
satisfies 
$$\left[t(u),\,t(v)\right]=0 \qquad \forall\,u,\,v\in\,{\mathbb C}.$$
This result does not depend on any properties of $r(u,v)$ other than (\ref{comm1},\ref{comm2}), as well as (\ref{comm}) being compatible with the Jacobi identity.
It can be shown that a {\it sufficient} condition for the Jacobi identity to hold is the imposition that the non-standard classical Yang-Baxter equation (CYBE)
\begin{align}
\left[r_{12}(u,v),\,r_{23}(v,w)\right]-\left[r_{21}(v,u),\,r_{13}(u,w)\right]+\left[r_{13}(u,w),\,r_{23}(v,w)\right] =0
\label{cybe}
\end{align}
holds. Recall that for the complex vector space ${\mathbb C}^m$,  a Lie algebra representation $\pi:{\mathcal G}\rightarrow {\rm End}({\mathbb C}^m)$ is required to preserve the commutator in the sense
\begin{align}
\pi\left(\left[T^j_k(u),\,T^p_l(v)\right]\right)&=\left[\pi(T^j_k(u)),\,\pi(T^p_l(v))\right].
\label{rep}
\end{align} 
The imposition of the CYBE also provides the notion of an {\it adjoint} representation $\pi_{\rm ad}:{\mathcal G}\rightarrow {\rm End}({\mathbb C}^n)$. The identification
\begin{align}
\pi_{\rm ad}(T^{j}_{k}(u))=\sum_{l,p=1}^n r^{jl}_{kp}(u,v)\otimes e^p_l
\label{adjoint}
\end{align}
satisfies (\ref{rep}) as a consequence of (\ref{comm}) and (\ref{cybe}).

The above construction is non-standard in the sense that there is no assumption that the solution of (\ref{cybe}) possesses the {\it skew-symmetry} property
\begin{align*}
r(u,v)=-r(v,u).
\end{align*}
Consequently such solutions lie outside the Belavin-Drinfel'd classification of solutions \cite{bd82}. The skew-symmetry property arises naturally in cases where a solution of (\ref{cybe}) is obtained from a solution $R(u,v)\in{\rm End}({\mathbb C}^n\otimes {\mathbb C}^n)$ of the Yang-Baxter equation
\begin{align*}
R_{12}(u,v)R_{13}(u,w)R_{23}(v,w)=R_{23}(v,w)R_{13}(u,w)R_{12}(u,v)
\end{align*}
with the unitarity condition 
\begin{align}
R_{12}(u,v)R_{21}(v,u)=I\otimes I,
\label{unitarity}
\end{align}
where $I$ is the identity operator. If $R(u,v)$ depends on some parameter $\eta$, and admits an expansion $R(u,v)\sim I\otimes I+\eta \,{\mathfrak r}(u,v)+ O(\eta^2)$ as $\eta\rightarrow 0$,
 then ${\mathfrak r}(u,v)$ satisfies 
\begin{align}
\left[{\mathfrak r}_{12}(u,v),\,{\mathfrak r}_{23}(v,w)\right]+\left[{\mathfrak r}_{12}(u,v),\,{\mathfrak r}_{13}(u,w)\right]
+\left[{\mathfrak r}_{13}(u,w),\,{\mathfrak r}_{23}(v,w)\right] =0
\label{sscybe}
\end{align} 
while it follows from (\ref{unitarity}) that 
\begin{align}
{\mathfrak r}_{12}(u,v)+{\mathfrak r}_{21}(v,u)=0.
\label{cunitarity}
\end{align}
It is then seen that (\ref{sscybe}) and (\ref{cunitarity}) combined are equivalent to (\ref{cybe}). 

As an example, let $P: {\mathbb C}^n \otimes {\mathbb C}^n \rightarrow {\mathbb C}^n \otimes {\mathbb C}^n $ denote the permutation operator with the action
\begin{align*}
P(x \otimes y)=(y \otimes x) \qquad \forall \, x,\,y\in\, {\mathbb C}^n.
\end{align*}
A simple solution of (\ref{cybe}) is 
\begin{align}
r(u,v)=\frac{1}{u-v}P.
\label{rat}
\end{align}
This solution has the property (\ref{cunitarity}), so it also provides a solution of (\ref{sscybe}). 

Hereafter set $n=2$. One family of solutions of (\ref{cybe}) which does not generally possess the skew-symmetry property 
(\ref{cunitarity}),  but does satisfy (\ref{comm1},\ref{comm2}), was presented by Skrypnyk \cite{s09}. This solution can be represented in matrix form as
\begin{align} 
r(u,v)
&= \left(\begin{array}{ccccc} \displaystyle
\frac{u^2+v^2}{u^2-v^2}+c &0&|&0&0 \\
0&0 &|&\displaystyle\frac{2uv}{u^2-v^2}&0 \\ 
-&-&~&-&- \\ 
0&\displaystyle\frac{2uv}{u^2-v^2}&|&0 &0 \\
0&0&|&0&\displaystyle\frac{u^2+v^2}{u^2-v^2}+c  \end{array} \right)  
\label{nonss}
\end{align}
where $c\in\,{\mathbb C}$ is arbitrary. The above solution is skew-symmetric only for the particular choice $c=0$. In this instance the algebra $\mathcal G$ coincides with the trigonometric/hyperbolic $sl(2)$ Gaudin algebra \cite{osdr05,vdv14}. The primary interest here is in the case $c=1$ which, as will be explained below, gives rise to an exotic symmetry. For this choice 
\begin{align} 
r(u,v)&= \left(\begin{array}{ccccc} 
\displaystyle\frac{2u^2}{u^2-v^2} &0&|&0&0 \\
0&0 &|&\displaystyle\frac{2uv}{u^2-v^2}&0 \\ 
-&-&~&-&- \\ 
0&\displaystyle\frac{2uv}{u^2-v^2}&|&0 &0 \\
0&0&|&0&\displaystyle\frac{2u^2}{u^2-v^2}  \end{array} \right).
\label{exo} 
\end{align}
This solution may also be expressed as
\begin{align*}
r(u,v)=\left(\frac{u}{u-v}+\frac{u}{u+v}\sigma_1 \sigma_2\right)P 
\end{align*}
where $\sigma={\rm diag}(1,-1)$. This form is convenient for calculations that follow in Subsect. 2.1.

\subsection{Symmetries}
\noindent
Symmetries of solutions of (\ref{cybe}) may be characterised in terms of the one-dimensional representations $\pi_0$ of their associated Lie algebra ${\mathcal G}$, i.e. $\pi_0:{\mathcal G}\rightarrow {\mathbb C}$. From (\ref{rep}) it is seen that for any one-dimensional representation 
\begin{align*}
\pi_0\left(\left[T^j_k(u),\,T^p_l(v)\right]\right)&=0.
\end{align*}
Equivalently, a one-dimensional representation may be expressed as a mapping $\rho(\T(u))=\J(u)\in {\rm End}({\mathbb C}^2)$
satisfying
\begin{align}
0=\left[\J_2(v),\,r_{12}(u,v)\right]-\left[\J_1(u),\,r_{21}(v,u)   \right].
\label{1d}
\end{align}
In \cite{s14} such a solution is termed a {\it generalised shift element}.

For the case of the solution (\ref{rat}), it is seen that all constant elements of ${\rm End}({\mathbb C}^2)$ satisfy (\ref{1d}). It is therefore natural to identify the symmetry of (\ref{rat}) as $gl(2)$. On the other hand for generic values of $c$ in (\ref{nonss}), solutions of (\ref{1d}) are given by constant diagonal elements of ${\rm End}({\mathbb C}^2)$. In this instance the symmetry of (\ref{nonss}) is identified as $u(1)$. For the specific choice $c=1$ in (\ref{nonss}) leading to (\ref{exo}), an exotic symmetry emerges. Setting $\J(u)=\D+u\OO$ where $\D$ is diagonal and $\OO$ is off-diagonal it is found that (\ref{1d}) is satisfied by $\J(u)$. First
\begin{align*}
\left[\J_2(v),\,r_{12}(u,v)\right]&-\left[\J_1(u),\,r_{21}(v,u)  \right] \\
&=u(\D_2+v\OO_2)\left(\frac{1}{u-v}+\frac{1}{u+v}\sigma_1 \sigma_2\right)P 
\\
&\qquad
-u\left(\frac{1}{u-v}+\frac{1}{u+v}\sigma_1 \sigma_2\right)P(\D_2+v\OO_2) \\
&\qquad-v(\D_1+u\OO_1)\left(\frac{1}{v-u}+\frac{1}{u+v}\sigma_1 \sigma_2\right)P \\
& \qquad +v\left(\frac{1}{v-u}+\frac{1}{u+v}\sigma_1 \sigma_2\right)P(\D_1+u\OO_1) \\
&=(\D_2-\D_1)P+\sigma_1 \sigma_2(\D_2-\D_1)P 
\end{align*}
where simplification to the final expression has relied on the fact that $\sigma\OO=-\OO\sigma$.
Now every diagonal operator on ${\mathbb C}^2$ may be expressed as $\D=aI +b\sigma$ for some $a,b\in\,{\mathbb C}$. It then follows that 
\begin{align*}
\left[\J_2(v),\,r_{12}(u,v)\right]-\left[\J_1(u),\,r_{21}(v,u) \right] 
&=b(\sigma_2-\sigma_1)P+b\sigma_1\sigma_2(\sigma_2-\sigma_1)P \\
&=b(\sigma_2-\sigma_1)P+b(\sigma_1-\sigma_2)P \\
&=0.  
\end{align*}
Because this solution for $\J(u)$ has an explicit dependence on the parameter $u$, it is considered to be exotic compared to the $gl(2)$ symmetry of (\ref{rat}) and the $u(1)$ symmetry of (\ref{nonss}). For later use the notation $\overline{\mathcal{G}}$ will be adopted for the  Lie algebra associated with the case of the solution (\ref{exo}). 

\subsection{General representations}
\noindent
General representations of $\overline{{\mathcal G}}$ are first obtained by an evaluation homomorphism $\rho_z$ to the enveloping algebra $U(sl(2))$ of $sl(2)$. Adopting the convention for the $sl(2)$ generators $\{S^z,\,S^+,\,S^-\}$ to satisfy the commutation relations
\begin{align*}
[S^z,\,S^{\pm}]=\pm S^{\pm}, \qquad [S^+,\, S^-]=2S^z,
\end{align*}
$\rho_z$ has the action
\begin{align*}
\rho_z(T^1_1(u))&=\frac{u^2}{u^2-z^2}(I+2S^z),   \\  
\rho_z(T^1_2(u))&=\frac{uz}{u^2-z^2}S^+,    \\
\rho_z(T^2_1(u))&=\frac{uz}{u^2-z^2}S^-,  \\
\rho_z(T^2_2(u))&=\frac{2u^2}{u^2-z^2}(I-2S^z),
\end{align*}
preserving the commutation relations (\ref{comm}) for the solution (\ref{exo}). Every representation of $sl(2)$ then affords a 
a representation of $\G$. For the choice of the spin-1/2 representation of $sl(2)$ this coincides with the adjoint representation (\ref{adjoint}).  

A second evaluation homomorphism to the enveloping algebra $U(sl(2))$, denoted $\rho_{(0)}$, is deduced from the one-dimensional representation. Specifically,
\begin{align*}
\rho_{(0)}(T^1_1(u))&=\alpha I,   \\  
\rho_{(0)}(T^1_2(u))&=u\beta I,    \\
\rho_{(0)}(T^2_1(u))&=u\gamma I,  \\
\rho_{(0)}(T^2_2(u))&=\delta I,
\end{align*}
where $\alpha,\,\beta,\,\gamma,\,\delta\in\,{\mathbb C}$ are arbitrary, which again preserves the commutation relations (\ref{comm}) for the solution (\ref{exo}). Taking the tensor product of these evaluation homomorphisms provides evaluation homomorphisms acting on tensor product of copies of $U(sl(2))$. The general map
\begin{align*}
\rho^{\otimes L}=\rho_{(0)}\otimes \rho_{z_1}\otimes \rho_{z_2}\otimes ...\otimes \rho_{z_L}
\end{align*}
has the form 
\begin{align*}
\rho^{\otimes L}(T^1_1(u))&=\alpha I+\sum_{j=1}^L\frac{u^2}{u^2-z_j^2}(I+2S_j^z),   \\  
\rho^{\otimes L}(T^1_2(u))&=u\beta I+\sum_{j=1}^L\frac{2uz_j}{u^2-z_j^2}S_j^+,    \\
\rho^{\otimes L}(T^2_1(u))&=u\gamma I+\sum_{j=1}^L\frac{2uz_j}{u^2-z_j^2}S_j^-,  \\
\rho^{\otimes L}(T^2_2(u))&=\delta I+\sum_{j=1}^L\frac{u^2}{u^2-z_j^2}(I-2S_j^z).
\end{align*}
It then follows that $\tau(u)=\rho^{\otimes L}(t(u))\in\,U(sl(2))^{\otimes L}$ satisfying
\begin{align*}
[\tau(u),\,\tau(v)]=0, \qquad \forall\,u,\,v\in\,{\mathbb C}
\end{align*}
is a transfer matrix expressed in terms of spin operators.

Hereafter set $\delta=-\alpha$, and $\gamma=\beta$ with $\alpha,\,\beta\in\,{\mathbb R}$, giving 
\begin{align*}
\tau(u)&=\left(\alpha I+\sum_{j=1}^L\frac{u^2}{u^2-z_j^2}(I+2S_j^z)\right)^2+\left(-\alpha I+\sum_{j=1}^L\frac{u^2}{u^2-z_j^2}(I-2S_j^z)\right)^2   \\  
&\qquad+\left(u\beta I+\sum_{j=1}^L\frac{2uz_j}{u^2-z_j^2}S_j^+\right)
\left(u\beta I+\sum_{k=1}^L\frac{2uz_k}{u^2-z_k^2}S_k^-\right) \\
&\qquad +\left(u\beta I+\sum_{j=1}^L\frac{2uz_j}{u^2-z_j^2}S_j^-\right)\left(u\beta I+\sum_{k=1}^L\frac{2uz_k}{u^2-z_k^2}S_k^+\right) \\
&= 2(\alpha^2+\beta^2u^2)I  + 2u^4\left(\sum_{j=1}^L\frac{1}{u^2-z_j^2}\right)^2I    
\\
&\qquad\qquad\qquad\qquad 
+4u^2\sum_{j=1}^L\frac{z_j^2}{(u^2-z_j^2)^2}C_j+4\sum_{j=1}^L\frac{u^2}{u^2-z_j^2}\T_j.
\end{align*}
where $C=2(S^z)^2+S^+S^-+S^-S^+$ is the $sl(2)$ Casimir element and 
\begin{align}
\T_j=2(S_j^z)^2+&2\alpha S_j^z+\beta z_j (S_j^++S_j^-) \nonumber \\ 
+&\sum_{k\neq j}^L\left(\frac{4z^2_j}{z_j^2-z_k^2}S_j^zS_k^z+\frac{2z_jz_k}{z_j^2-z_k^2}(S_j^+S_k^-+S_j^-S_k^+)\right).
\label{ttees}
\end{align}
With the restriction $z_j\in\,{\mathbb R}, \,j=1,...,L$, which is imposed from now on, the $\{\T_j:j=1,...,L\}$ form a set of mutually commuting, self-adjoint operators.  Consequently they are simultaneously diagonalisable.
It is seen that the union of this set, the identity operator, and the set of Casimir elements $\{C_j:j=1,...,L\}$ is the full set of functionally independent operators that can be extracted from the transfer matrix.

Matrix forms of transfer matrix are obtained by choosing a representation of $sl(2)$ for each copy in the tensor product 
$U(sl(2))^{\otimes L}$. It should be noted that these representations need not be isomorphic, and throughout the parameters 
$\alpha,\,\beta, z_1,\,z_2,\,...,z_L$ are arbitrary real parameters. This provides a range of potential applications for constructing integrable models.

\section{Integrable, multi-species, boson tunneling model}

\noindent
Having identified the exotic symmetry of the solution (\ref{exo}), and outlined the construction of representations of $\G$, attention turns towards an application through the construction of an integrable, multi-species, boson tunneling model. The Hamiltonian is defined as
\begin{align}
H&=2U\sum_{j=1}^L \T_j \nonumber \\
&= 4U\left(\sum_{j=1}^L S_j^z\right)^2+4U\alpha\sum_{j=1}^L S_j^z +2{U\beta}\sum_{j=1}^L z_j(S_j^++S_j^-).
\label{sum}
\end{align}
Next the spin operators are represented in terms of boson operators through the Jordan-Schwinger map
\begin{align*}
S^z_j&=\frac{1}{2}(N_{j,A}-N_{j,B}), \\
S^+_j&=b^\dagger_{j,A}b_{j,B},  \\
S^-_j&=b^\dagger_{j,B}b_{j,A}
\end{align*}
where the spin $s_j$ of each representation is determined by $2s_j=N_j$.
This maps (\ref{sum}) to (\ref{ham}) with the identification $\mu=2U\alpha$ and ${\mathcal E}_j=2U\beta z_j$. 
For simplicity it will be assumed hereafter that the $z_j$, and consequently the ${\mathcal E}_j$, are distinct parameters. 

Since the operators (\ref{ttees}) commute with (\ref{sum}), application of the Jordan-Schwinger map to the $\T_j$ yields a set of $L$ operators which necessarily commute with (\ref{ham}). Along with the $N_j$, which also commute with (\ref{ham}), the construction for the model is one which establishes that it is integrable since the number of independent conserved operators is the same as the number of degrees of freedom. Closely related to the property of integrability is the notion of an exact solution. The next step is to derive an explicit form for such a solution.  

\subsection{Bethe Ansatz solution}
\noindent
Below it will be shown that the eigenvalues $\Lambda_j$ of the conserved operators (\ref{ttees})  have the form 
\begin{align}
\Lambda_j&=2s_j^2+2\alpha s_j+\sum_{k\neq j}^L\frac{4s_js_k z_j^2}{z_j^2-z_k^2} -2\sum_{n=1}^N\frac{s_j z_j^2}{z_j^2-v_n} 
\label{eigs}
\end{align}
such that the parameters $\{v_n:l=1,...,N\}$, known as the Bethe roots, satisfy the Bethe Ansatz equations
\begin{align}
\alpha-1+\sum_{l=1}^L\frac{2v_n s_l}{v_n-z_l^2}&=\sum_{m\neq n}^N\frac{2v_n}{v_n-v_m}
+\frac{\displaystyle \beta^2\prod_{l=1}^L(v_n-z_l^2)^{2s_l}} {\displaystyle 4\prod_{m\neq n}^N(v_n-v_m)}.
\label{bae}
\end{align}
Consequently, the energy eigenvalues $E$ for the Hamiltonian (\ref{ham}) have the form
\begin{align*}
E&=2U\sum_{j=1}^L \Lambda_j \\
&=UN^2+\mu N-4U\sum_{j=1}^L\sum_{n=1}^N \frac{ s_j z_j^2}{z_j^2-v_n}.
\end{align*} 

Let the representation space associated with highest weight $s$ be denoted $V(s)$, such that for a system with $N_j$ bosons of each species $j$ the Hilbert space is 
\begin{align*}
V(N_1/2)\otimes V(N_2/2)\otimes ... \otimes V(N_L/2)
\end{align*}  
The tensor product state consisting of all the highest weight states is not an eigenstate of the transfer matrix,  
except in the case when $\beta=0$. Consequently, the algebraic Bethe Ansatz method of solution for calculating eigenvalue expressions of the operators 
(\ref{tees}) cannot be easily applied due to the absence of an apparent reference state.  
Here a different approach is taken, based on operator identities in the first instance for a specific case. Using the method of proof by induction, the result will be extended to the general case by examining tensor product decompositions.    

\subsection{$N$-species case}
\noindent 
First the case where $N_j=1$ for all $j=1,...,L$, so $N=L$, will be considered. The quasi-classical limit of the {\it Off-diagonal} Bethe Ansatz approach \cite{wycs15} was utilised in \cite{lil16} to obtain the solution in this instance. While the physical interpretation of the $p+ip$ pairing Hamiltonian adopted in \cite{lil16} is a completely different context to the one considered here, at the mathematical level it is entirely equivalent and can be seen as the case where all copies of $sl(2)$ algebras are represented by spin 1/2.  It was found that the eigenvalues $\lambda_j$ are given by (\ref{eigs}) subject to the Bethe Ansatz equations (\ref{bae}) with the specialisation $s_j=1/2$ for all $j=1,...,L$. In \cite{cdv16} a method which generalises the algebraic Bethe Ansatz was used to confirm the result.

The approaches of \cite{cdv16,lil16} are quite technical. Below a different approach will be taken, following \cite{l16}, which reproduces the equations in a streamlined manner. To simplify calculations, first define the operators 
\begin{align}
T_j=2\alpha S_j^z&+\beta z_j (S_j^++S_j^-) \nonumber \\ 
&+\sum_{k\neq j}^L\left(\frac{z^2_j}{z_j^2-z_k^2}(4S_j^zS_k^z-I)+\frac{2z_jz_k}{z_j^2-z_k^2}(S_j^+S_k^-+S_j^-S_k^+)\right)
\label{tees}
\end{align}
which, in the spin-1/2 case, only differ from (\ref{ttees}) by constant terms.
The starting point uses a result given in \cite{cdv16} that the operators $T_j$ satisfy the 
quadratic identities
\begin{align*}
T_j^2=\alpha^2+\beta^2z_j^2-2z_j^2\sum_{k\neq j}^L\frac{1}{z_j^2-z_k^2}(T_j-T_k).
\end{align*}
Since these are operator identities it follows that the eigenvalues $\lambda_j$ of the operators $T_j$ satisfy analogous relations:
\begin{align}
\lambda_j^2=\alpha^2+\beta^2z_j^2-2z_j^2\sum_{k\neq j}^L\frac{\lambda_j-\lambda_k}{z_j^2-z_k^2}.
\label{quad}
\end{align}
Define $Q(u)$ to be a polynomial of order $L$ satisfying 
\begin{align*}
2z_j^2Q'(z_j^2)+(\lambda_j-\alpha)Q(z_j^2)=0, \qquad j=1,...,L.
\end{align*}
Since this is a homogeneous linear system of $L$ equations for the $L+1$ coefficients 
of the polynomial $Q(u)$, there exists a non-trivial solution. Set 
\begin{align*}
Q(u)=\prod_{k=1}^L(u-v_k),
\end{align*}
which assumes that the roots of $Q(u)$ do not have multiplicities. The validity of this assumption will be reviewed later.
Now, provided $Q(z_j^2)\neq 0$ for all $j=1,...,L$, then 
\begin{align}
\lambda_j^2
&= \alpha^2-4\alpha z_j^2\frac{Q'(z_j^2)}{Q(z_j^2)}+4z_j^4\left( \frac{Q'(z^2_j)}{Q(z^2_j)}\right)^2 \nonumber \\
&= \alpha^2-4\alpha z_j^2\frac{Q'(z^2_j)}{Q(z^2_j)}+4z_j^4\sum_{k=1}^L\frac{1}{(z_j^2-v_k)^2} +4z_j^4\frac{Q''(z^2_j)}{Q(z^2_j)}.  
\label{sq2}
\end{align}

On the other hand from (\ref{quad})
\begin{align}\lambda_j^2&=\alpha^2+\beta^2z_j^2
-2z_j^2\sum_{k\neq j}^L\frac{\lambda_j-\lambda_k}{z^2_j-z^2_k}\nonumber\\
&=\alpha^2+\beta^2z_j^2+4z_j^2\sum_{l=1}^L\sum_{k\neq j}^L\frac{v_l}{(z^2_k-v_l)(v_l-z^2_j)} \nonumber\\
&=\alpha^2+\beta^2 z_j^2+4z_j^2\sum_{l=1}^L\frac{v_l}{z^2_j-v_l}\frac{P'(v_l)}{P(v_l)} \nonumber\\
&\qquad\quad +4z_j^2\sum_{l=1}^L\frac{z_j^2}{(v_l-z^2_j)^2} 
+4z_j^2\sum_{l=1}^L\frac{1}{v_l-z_j^2}\nonumber\\
&=\alpha^2+\beta^2 z_j^2+\sum_{l=1}^L\frac{4z_j^2 v_l}{z^2_j-v_l}\frac{P'(v_l)}{P(v_l)} +\sum_{l=1}^L\frac{4z_j^4}{(v_l-z^2_j)^2} 
-4z_j^2\frac{Q'(z_j^2)}{Q(z_j^2)}\label{sq1} 
\end{align}
where
\begin{align*}
P(u)=\prod_{j=1}^L(u-z_j^2).
\end{align*}
%
For equality of the expressions (\ref{sq2}) and (\ref{sq1}) it is required that
\begin{align}
\frac{\beta^2}{4} +\sum_{l=1}^L\frac{v_l}{z^2_j-v_l}\frac{P'(v_l)}{P(v_l)}
=(1-\alpha) \frac{Q'(z^2_j)}{Q(z^2_j)}+z_j^2\frac{Q''(z^2_j)}{Q(z^2_j)}. 
\label{zeros}
\end{align}
Set
\begin{align}
S(u)=uQ''(u)+(1-\alpha) Q'(u)-\left(\frac{\beta^2}{4}+\sum_{j=1}^L\frac{v_j}{u-v_j}\frac{P'(v_l)}{P(v_l)}\right)Q(u)
\label{ess}
\end{align}
which is a polynomial of order $L$. It follows from (\ref{zeros}) that
\begin{align*}
S(z_j^2)=0,\qquad j=1,..., L
\end{align*}
which, along with the consideration of the asymptotic behaviour of (\ref{ess}) as $u\rightarrow\infty$, establishes that
\begin{align}
S(u)&=-\frac{\beta^2}{4}P(u).
\label{essp}
\end{align}
Evaluating $S(v_n)$ through (\ref{ess}) and (\ref{essp}) and equating these expressions then yields the Bethe Ansatz equations 
\begin{align}
\alpha-1+v_n\frac{P'(v_n)}{P(v_n)}=v_n\frac{Q''(v_n)}{Q'(v_n)}+\frac{\beta^2}{4}\frac{P(v_n)}{Q'(v_n)},
\qquad n=1,...,L
\label{poly}
\end{align}
which are equivalent to (\ref{bae}). It can be verified that (\ref{eigs}) holds.

It needs addressing at this point that the assumptions  that the roots of $Q(u)$ do not have multiplicities, and that $Q(z_j^2)\neq 0$, are not always valid. In the $\beta=0$ limit numerical solutions of the  Bethe Ansatz equations confirm that both of these situations occur \cite{dilsz10,ilsz09,lmm15,rdo10}. Two Bethe roots may coalesce at a point $z_j^2$ through which they transition between being real-valued and a complex-conjugate pair. Moreover in this limit some of the $L$ Bethe roots must be seen as being infinite-valued, and there are well-identified points at which several additional roots collapse to be zero-valued or infinite-valued. All of these instances may be interpreted as singular solutions of (\ref{bae}), but can be simply regularised by expressing (\ref{poly}) in a polynomial form
\begin{align*}
(\alpha-1)P(v_n)Q'(v_n)+v_n{P'(v_n)}{Q'(v_n)}=v_n{Q''(v_n)}{P(v_n)}+\frac{\beta^2}{4}{P(v_n)^2}.
\end{align*}
So, while the above {\it derivation} of the Bethe Ansatz equations breaks down when the roots of $Q(u)$ have multiplicities, or $Q(z_j^2)=0$, indicators from available numerical studies \cite{dilsz10,ilsz09,lmm15,rdo10} suggest that the equations are nonetheless valid.    

\subsection{The general case}
\noindent 
The validity of the formulae (\ref{eigs},\ref{bae}) in general follows by an inductive argument on each of the species number labels $N_j$, starting with the case $N_j=1$ for all $j=1,...,N$ which has been established above. 
Every finite-dimensional irreducible $sl(2)$ module is contained in some tensor product of spin-1/2 modules, and in particular
\begin{align*}
V(N_1/2)\otimes ...V(N_L/2) \subseteq V(1/2)^{\otimes N}
\end{align*}
where $N=N_1+...+N_L$ as before.
Now consider a system of $M\geq L$ species, for which it is determined that  
for $j\neq M-1,\,M$
\begin{align}
\lim_{z_M\rightarrow z_{M-1}}\T_j=&2(S_j^z)^2+2\alpha S_j^z+\beta z_j (S_j^++S_j^-) \nonumber \\ 
&\quad+\sum_{k\neq j}^{M-2}\left(\frac{4z^2_j}{z_j^2-z_k^2}S_j^zS_k^z+\frac{2z_jz_k}{z_j^2-z_k^2}(S_j^+S_k^-+S_j^-S_k^+)\right)
\nonumber \\
&\qquad +\frac{4z^2_j}{z_j^2-z_{M-1}^2}S_j^z(S_{M-1}^z+S_M^z) \nonumber \\
&\qquad+\frac{2z_jz_{M-1}}{z_j^2-z_{M-1}^2}(S_j^+(S_{M-1}^-+S_M^-)
+S_j^-(S_{M-1}^++S_M^+))
\label{mod1}
\end{align}
while
\begin{align}
\lim_{z_M\rightarrow z_{M-1}}(\T_{M-1}+\T_M)&=2(S_{M-1}^z+S_M^z)^2+2\alpha (S_{M-1}^z+S_M^z) \nonumber \\
&\qquad+\beta z_j (S_{M-1}^++S_M^++S_{M-1}^-+S_M^-) \nonumber \\ 
&\qquad+\sum_{k=1}^{M-2}\frac{2z_{M-1}z_k}{z_{M-1}^2-z_k^2}(S_{M-1}^++S_M^+)S_k^-
\nonumber  \\
&\qquad+\sum_{k=1}^{M-2}\frac{2z_{M-1}z_k}{z_{M-1}^2-z_k^2}(S_{M-1}^-+S_M^-)S_k^+
\nonumber  \\
&\qquad+\sum_{k=1}^{M-2}\frac{4z^2_{M-1}}{z_{M-1}^2-z_k^2}(S_{M-1}^z+S_M^z)S_k^z .
\label{mod2}
\end{align}
Noting the result  
\begin{align*}
V(s_{M-1}+s_M)\subseteq V(s_{M-1})\otimes V(s_M),
\end{align*}
let $\P_M$ denote the projection onto the component $V(s_{M-1}+s_M)$. 
The above expressions (\ref{mod1},\ref{mod2}) show that the general form of the conserved operators (\ref{ttees}) is preserved via
\begin{align*}
\tilde{\T}_j&=\lim_{z_M\rightarrow z_{M-1}}\P_M\T_j \P_M, \qquad j=1,...,M-2, \\
\tilde{\T}_{M-1}&= \lim_{z_M\rightarrow z_{M-1}}\P_M(\T_{M-1}+\T_M)\P_M
\end{align*}
whereby the tilded notation refers to a new system of $N$ particles of $M-1$ species, such that $\tilde{N}_{M-1}=N_{M-1}+N_M$ and $\tilde{N}_j=N_j$ otherwise.   
It should be noted that a projection onto any other irreducible component of the tensor product $V(s_{M-1})\otimes V(s_M)$ gives rise to a system where the number of particles is strictly less that $N$. This is not an appropriate step for establishing the desired result, since it requires a change in the number of Bethe roots.

Suppose that (\ref{eigs},\ref{bae}) hold for the $M$-species system. Letting $\tilde{\Lambda}_j$ denote the eigenvalues for $\tilde{\T}_j$, it is required to show that (\ref{eigs},\ref{bae}) hold for the new system of $M-1$ species with the tilded notation. For (\ref{bae}) the result follows simply from the relations
\begin{align*}
\lim_{z_M\rightarrow z_{M-1}}\sum_{l=1}^{M}\frac{2s_l}{v_n-z_l^2}
&=
\frac{2(s_{M-1}+s_{M})}{v_n-z_{M-1}^2}+\sum_{l=1}^{M-2}\frac{2s_l}{v_n-z_l^2}, \\
\lim_{z_M\rightarrow z_{M-1}}\prod_{l=1}^M(v_n-z_l^2)^{2s_l}
&=(v_n-z_{M-1})^{2s_{M-1}+2s_M}{\prod_{l=1}^{M-2}(v_n-z_l^2)^{2s_l}}.
\end{align*} 
In a similar manner it follows that ({\ref{eigs}) holds with the observations
\begin{align*}
\tilde{\Lambda}_j&= \lim_{z_M\rightarrow z_{M-1}}\Lambda_j, \\
\tilde{\Lambda}_{M-1}&= \lim_{z_M\rightarrow z_{M-1}}(\Lambda_{M-1}+\Lambda_M).
\end{align*} 
Finally note that at each step the labels $N_j$ can be reordered without loss of generality, such that limits (\ref{mod1},\ref{mod2}) can be taken. It therefore follows by induction on each $N_j=2s_j$ that (\ref{eigs},\ref{bae}) hold in general. 

\section{Conclusion}

\noindent
This work first identified a particular solution of a non-standard form of the classical Yang-Baxter equation. This solution is unusual in that it admits a symmetry which is more general than $u(1)$-symmetry, but less general than $gl(2)$-symmetry. The solution was used for the construction of an integrable, multi-species, boson tunneling model, and the Bethe Ansatz equations characterising the spectrum were derived. 

Having obtained the exact solution, there are several avenues for ongoing work to take place. One promising course is to adapt integral equation techniques \cite{dilsz10,rdo10} to investigate ground-state properties in the limit of large particle numbers. This will be the subject of a future study.

\section*{Acknowledgments}

\noindent
This work was supported by the Australian Research Council through Discovery Project DP150101294. I thank Taras Skrypnyk for helpful comments.


\end{document}